\documentclass[pra,aps,twocolumn,showpacs,preprintnumbers,amsmath,amssymb,floatfix]{revtex4}
\usepackage{epsfig}
\usepackage{graphics}
\usepackage{dcolumn}
\usepackage{bm}
\usepackage{mathrsfs}
\newcommand{\bq}{\begin{eqnarray}}
\newcommand{\eq}{\end{eqnarray}}

\begin{document}

\title{Quantum dynamics of Raman-coupled Bose-Einstein condensates using Laguerre-Gaussian beams}

\author{Rina Kanamoto, Ewan M. Wright, and Pierre Meystre}
\affiliation{
Department of Physics and College of Optical Sciences, The University of Arizona, Tucson, Arizona 85721, USA
}
\date{\today}

\begin{abstract}
We investigate the quantum dynamics of Raman-coupled Bose-Einstein 
condensates driven by laser beams that carry orbital angular 
momentum. By adiabatically eliminating the excited atomic state we 
obtain an effective two-state Hamiltonian for the coupled 
condensates, and quantization of the matter-wave fields results in 
collapse and revivals in the quantum dynamics. We show that the revival 
period depends on whether 
the initial nonrotating condensate displays broken U(1) 
symmetry or not, and that the difference may be detected by 
measuring the motion of quantized vortices that are nested in the 
density profile of the Raman-coupled condensates. We further study 
the steady-state population transfer using a linear sweep of the 
two-photon detuning, by which the atomic population is coherently 
transferred from an initial nonrotating state to the final vortex state.
\end{abstract}
\maketitle

\section{Introduction}

Controlling and probing of the quantum state of cold atoms or 
molecules are opening up new frontiers for exploring the 
fundamentals of quantum mechanics. At the ultracold temperatures 
characteristic of Bose-Einstein condensates (BECs) matter reveals 
its wave character, and the quantum statistics of the atoms or 
molecules play an important role. By controlling both the atomic 
collisions and the properties of laser fields applied to a BEC, 
one may generate tailor made quantum states of matter. For 
example, photons and atoms can interchange mechanical 
momentum, and atoms can therefore be prepared 
into a specific center-of-mass momentum state by the appropriate 
choice of spatial light field profile~\cite{97AR,86ADBC}. 
The light field may also be used for the detection of the atomic quantum 
state~\cite{cqed}.

A commonly applied notion for atomic BECs is that of Bose broken 
symmetry (BBS)~\cite{Yang,And84,Griffin,LS91}, or broken U(1) 
symmetry, whereby the state vector for the BEC is described via a 
wave packet of states of varying atom number $N$.  In this scheme 
the fact that the atom number is not fixed is traded off for the 
fact that the macroscopic wave function for the BEC, obtained as 
the expectation value of the field annihilation operator, acquires 
a definite phase, hence the name broken U(1) 
symmetry.  The BBS approach is typified by the use of a 
coherent-state description for the state of a BEC.  In contrast, a 
number conserving (NC) approach is also possible, but in that case 
the phase of the corresponding macroscopic wave function has no 
physical significance, and U(1) symmetry is 
preserved. In the absence of atomic collisions, 
it is known that whether or not the U(1) symmetry is broken does 
not affect any physical observables, and whether one describes the 
state of the BEC as a coherent state or a number state is largely 
a matter of calculational convenience~\cite{And84,LS91}.  Furthermore, 
for the case involving the interference between two BECs, even for 
an initial state with a fixed number of atoms, quantum coherence is 
built up by the measurement process which leads to uncertainty in 
the atom number, and the resulting interference pattern is 
indistinguishable from that which would have resulted from a 
coherent state description ~\cite{96JY,97CD}.

However, this equivalence between the BBS and NC approaches holds 
only when atomic collisions can be neglected. If one takes atomic 
collisions into account differences between the two approaches 
appear in the collapse and revival times that can occur in the 
interference between BECs with different spatial modes, for 
example in the interference between condensates~\cite{97WCWTW,Walls2}, 
a double-well system~\cite{97MCWW}, or in Ramsey 
fringe-type experiments~\cite{01S}. Intuitively, the revival 
period difference arises from the fact that the allowed number 
difference $\Delta N$ between the two BECs is different for the 
two cases: In the NC approach $\Delta N$ is always an even 
integer, whereas in the coherent state approach based on BBS 
$\Delta N$ can be an arbitrary integer.  Since atomic collisions 
lead to matter wave phase shifts proportional to the atom number, 
the interference between two BECs, which depends on $\Delta N$, is 
markedly different depending on whether U(1) symmetry is broken 
or not, and leads to a factor of two difference in the revival 
period between the two approaches. This factor of two difference 
has recently been observed for the first time in an experiment 
that looks at the interference between atoms released from a 
lattice of double well systems~\cite{07MD}: Depending on the 
preparation conditions, the atoms trapped in each individual 
double well in the lattice 
may be best described as a number state or a coherent state, and 
this leads to a factor of two difference in the revival time in 
the interference pattern that arises when the atoms are released 
and allowed to interfere.

Collapse and revivals have previously been investigated experimentally in the 
Jaynes-Cummings model of a two-level atom interacting with a 
quantized laser field~\cite{87RWK}, 
wave packet dynamics of Rydberg atoms~\cite{Rydberg}, matter waves in an optical 
lattice~\cite{02GMHB}, and most recently for a matter wave in a 
lattice of double wells~\cite{07MD}. Nevertheless, the use of 
collapse and revivals to test the fundamental notion of U(1) 
broken-symmetry as applied to a {\it single} BEC via the period of 
the collapse and revivals has not been realized experimentally so 
far.  Walls and co-workers ~\cite{97WCWTW,Walls2} showed that this could be realized by 
coupling an initial BEC with a given mode structure to a second 
BEC mode, thereby realizing a two-mode system.  The 
period of the revivals of the system would then act as a test of 
whether or not the initial isolated BEC exhibited BBS or not. 
Search \cite{01S} has proposed precisely such an 
experiment based on a Ramsey fringe approach, but this has not 
been realized to date.

In this paper, we consider the case of Raman-coupled BECs using 
Laguerre-Gaussian (LG) beams~\cite{97MZW} as a potential testing ground for 
the BBS description of an initial BEC.  Because of the helical 
phase structure of the laser fields, orbital angular momentum 
(OAM) is transferred from the light to the atoms and results in a 
BEC in a coherent superposition of two components with distinct 
center-of-mass OAM~\cite{97MZW,98DCLZ,05KD}. (In another scheme for generating 
vortices a de-centered Gaussian beam is rotated around the center of a trapped 
BEC~\cite{MatAnd99}, but we do not analyze that case here.) 
By considering BECs 
with relatively small numbers of atoms, and quantizing the 
matter-wave fields, the system can be mapped onto a Hamiltonian 
that describes the dynamics of an ideal two-mode system. Then the 
granular nature of the matter-wave field becomes important, and we 
explore the collapse and revivals in the system both with and 
without broken U(1) symmetry. In particular, we show that the 
difference between the two approaches is directly visible in the 
motion of quantized vortices that appear in the density profile of the 
Raman-coupled condensates, the density profile being a relatively 
easy physical quantity to observe. To put our proposal in context, 
collapse and revival dynamics have previously been observed using 
the spatial density profile in a two-component $^{87}$Rb 
BEC~\cite{99JILA}, and the Raman vortex coupler using LG beams has 
now been realized experimentally~\cite{06NIST}, both of these 
experiments being performed for large atom numbers where the 
granular nature of the matter wave fields is not relevant.

This paper is organized as follows. In Sec.~\ref{formulation} we 
formulate the problem, and map the three-level $\Lambda$-type 
atomic system coupled with LG fields onto a two-mode Hamiltonian 
in the angular-momentum representation. In Sec.~\ref{results}, we 
focus on the collapse and revival dynamics for initial number 
states and coherent states for the BEC, and show that these may be 
measured using the spatial density profile of the Raman-coupled 
condensates as well as the atom statistics. 
Section~\ref{summary} summarizes the results in this paper, 
mentioning the time-dependent two-photon 
detuning for the stimulated Raman population transfer.

%
\section{Formulation of the problem}\label{formulation}

In this section we formulate the two-mode approximation for two 
Raman-coupled BECs and introduce the angular momentum 
representation used to solve the quantum dynamical system.

%
\subsection{Hamiltonian}

\begin{figure}[t]
\begin{center}
\epsfig{file=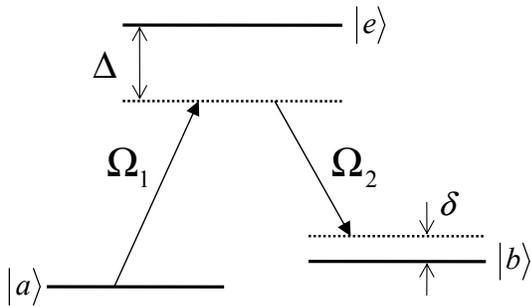,width=7cm}
\caption{
$\Lambda$ atomic level scheme: 
Three hyperfine states $|a\rangle, |b\rangle$,  and 
$|e\rangle$ are coupled with light fields of 
Rabi frequencies $\Omega_1$, $\Omega_2$, 
and laser frequencies $\omega_1$, $\omega_2$, 
respectively. 
The common detuning $\Delta$ is sufficiently large 
that the excited state $|e\rangle$ is not significantly populated.
}
\label{fig1}
\end{center}
\end{figure}

We consider a quantum-degenerate sample of ultracold bosonic atoms 
with three participating levels $|a\rangle, |b\rangle$, and 
$|e\rangle$ arranged in a $\Lambda$ configuration, see 
Fig.~\ref{fig1}. The sample is irradiated by a pair of laser 
fields of frequencies $\omega_{1,2}$, transitions between 
$|a\rangle$ and $|b\rangle$ being dipole forbidden. The system 
Hamiltonian $\hat{H}=\hat{H}_{\rm A}+\hat{H}_{\rm AF}$ is given by~\cite{Raman_Hamiltonian} 
\bq
\hat{H}_{\rm A}&=&
\sum_{j=a,b,e}\int d^3 r 
\hat{\Psi}_j^{\dagger}(\bm{r})\hat{H}_0^{(j)}\hat{\Psi}_j(\bm{r})\nonumber\\
+\ \frac{1}{2}\!\!\!\!\!\!&&\!\!\!\!\!\!\sum_{j,j'=a,b,e} \chi_{jj'}\!\!\int d^3 r
\hat{\Psi}_j^{\dagger}(\bm{r})\hat{\Psi}_{j'}^{\dagger}(\bm{r})
\hat{\Psi}_{j'}(\bm{r})\hat{\Psi}_j(\bm{r}),\label{HA}
\label{HamiltonianA0}\\
\hat{H}_{\rm AF}&=&-\hbar\int d^3 r
\left[
\hat{\Psi}_{e}^{\dagger}(\bm{r})
\hat{\Psi}_a(\bm{r})
\Omega_1^{(+)}(\bm{r})
e^{-i\omega_1 t}
+{\rm h.c.}\right.
\nonumber\\
&\ &\qquad \left. +\hat{\Psi}_{e}^{\dagger}(\bm{r})
\hat{\Psi}_b(\bm{r}) \Omega_2^{(+)}(\bm{r}) e^{-i\omega_2 t} +{\rm
h.c.} \right], \label{HamiltonianAF0}
\eq
where $\hat{\Psi}_j^\dagger,\hat{\Psi}_j$ are creation and annihilation 
field operators for bosonic atoms in the states $j=a,b,e$. Atomic 
collisions are included via the pseudopotential coefficients 
$\chi_{jj'}=4\pi\hbar^2 a_{jj'}/M$, where $a_{jj'}$ is the 
$s$-wave scattering length between two atoms in the states $j$ and 
$j'$, and $M$ is the atomic mass. The coupling of the atoms to the 
classical optical fields is described by the Rabi frequencies 
$\Omega^{(\pm )}_{1,2}$, where $\pm$ denotes the positive and 
negative frequency components. The atoms are further supposed to 
be confined in harmonic traps, so that the center-of-mass 
Hamiltonian for atoms in each state $j=a,b,e$ is given by
    \bq
    \hat{H}_0^{(j)}&=&-\frac{\hbar^2\nabla^2}{2M}+V_j(\bm{r}),\\
    V_j(\bm{r})&=&\frac{M \omega_{r,j}^2 (x^2+y^2)}{2}+\frac{M \omega_{z,j}^2 z^2}{2}, 
    \eq
with $\omega_{r,j}$ and $\omega_{z,j}$ the oscillator frequencies 
along the radial ($r$) and longitudinal ($z$) directions for the 
different atomic states.

If the detuning $\Delta$ from the excited state $|e\rangle$ 
is sufficiently large (see Fig. 1), the population of that state 
remains negligible and spontaneous emission may be neglected. 
The excited state can then be adiabatically eliminated as
    \bq
    \hat{\Psi}_e(\bm{r})\!=\!\frac{1}{\Delta}\!\left[\Omega_1^{(+)}\!(\bm{r})
    \hat{\Psi}_a(\bm{r})e^{-i\omega_1 t} \!+\!
    \Omega_2^{(+)}\!(\bm{r})\hat{\Psi}_b(\bm{r})e^{-i\omega_2t}\right]
    \eq
and the Hamiltonian for the atom-field interaction reduces to
    \bq\label{HAFae}
    \hat{H}_{\rm AF}=\frac{- 2 \hbar}{\Delta}\int d^3 r
    \left[ \Omega_1^{(+)}\Omega_2^{(-)}e^{i(\omega_2-\omega_1) t}
    \hat{\Psi}_b^{\dagger}\hat{\Psi}_a
    +{\rm  h.c.}\right],
    \eq
where we have omitted the diagonal terms due to the ac-stark shift, 
which create an effective potential for the atoms but do not affect the dynamics.
The atom-field interaction Hamiltonian $H_{\rm AF}$ in Eq.~(\ref{HAFae}) 
reflects the fact that if an atom makes a transition from state $|a\rangle$ to 
the excited state via absorption of a photon of frequency $\omega_1$, 
this will be rapidly followed by emission of a photon of frequency 
$\omega_2$ and a Raman transition to the state $|b\rangle$.

Bringing these results together, the system Hamiltonian 
$\hat{H}=\hat{H}_{\rm A}+\hat{H}_{\rm AF}$ is now given by
\bq
\hat{H}_{\rm A}\!\!&=&\!\!\!
\sum_{j=a,b}\int \! \!d^3 r \hat{\Psi}_j^{\dagger}\hat{H}_0^{(j)}\hat{\Psi}_j
\!+\!\!\!\sum_{j,j'=a,b} \!\!\frac{\chi_{jj'}}{2}\!\!\int \!\!d^3 r
\hat{\Psi}_j^{\dagger}\hat{\Psi}_{j'}^{\dagger}\hat{\Psi}_{j'}\hat{\Psi}_j,
\nonumber\\
\hat{H}_{\rm AF}\!\!&=&\!\!\frac{- 2 \hbar}{\Delta}\!\int d^3 r
    \left[\Omega_1^{(+)}\Omega_2^{(-)}e^{i(\omega_2-\omega_1)t}
    \hat{\Psi}_b^{\dagger}\hat{\Psi}_a+{\rm h.c.}  \right].
\label{HHH}
\eq
This Hamiltonian is the starting point for our analysis.

%
\subsection{Orbital angular momentum of light}

Light has two kinds of angular momentum: a spin-angular momentum 
associated with its polarization, and an orbital angular momentum 
(OAM) associated with mechanical rotation~\cite{03OAMreview}. The system described by 
the Hamiltonian~(\ref{HHH}) corresponds to a familiar coherent 
coupler between the atomic BECs in the states $|a\rangle$ and 
$|b\rangle$ if both light fields are plane waves or Gaussian 
beams, that is, beams that carry no OAM. Then the 
absorption-emission process transfers the atom from the state 
$|a\rangle$ to the state $|b\rangle$ that is shifted in momentum 
by the two-photon recoil. On the other hand, if one or both of 
those laser fields carries OAM, the light field can impart its OAM 
to the atoms, and hence angular-momentum transfer occurs 
associated with the coherent coupling \cite{VanEnk}. In this case, 
the two-photon recoil is associated with the difference of angular 
momenta of two light beams, and one realizes a Raman vortex 
coupler.  The Raman vortex coupler has previously been studied 
theoretically in the limit of large atom numbers where a treatment 
based on coupled Gross-Pitaevskii equations is applicable~\cite{97MZW}, 
and an experimental realization of this case has been reported \cite{06NIST}. 
Here we explore the opposite limit of low atom numbers, where the granular nature 
of the quantum matter wave fields becomes relevant.

A typical set of light fields that carry OAM are Laguerre-Gaussian beams 
whose mode functions at the beam waist $z=0$ are given by~\cite{03OAMreview}
\begin{eqnarray}
{\cal LG}_p^{l} (r,\varphi)\propto
(-1)^p\left(\frac{r}{w_0}\right)^{\!|l|} L_p^{|l|}\left(\frac{r^2}{w_0^2}\right)
e^{-r^2/2w_0^2}e^{il\varphi},
\end{eqnarray}
where $w_0$ is the focused spot size at $z=0$, $p$ is the radial mode number, 
$l$ the winding number, which describes the helical structure of the wave front, and
\begin{eqnarray}
L_p^{|l|}(r)=\sum_{j=0}^p  \frac{(|l|+p)!(-r)^j}{(p-j)!(|l|+j)!j!},
\end{eqnarray}
are Laguerre polynomials. By using LG modes for the Raman coupling between 
the matter wave fields in the Hamiltonian~(\ref{HHH}) we can therefore create 
a matter-wave vortex even though the initial state is nonrotating. 
Henceforth, we consider only the LG modes with $p=0$, and we thus need only 
consider one quantum number $l$, to specify the OAM.

%
\subsection{Two-mode Hamiltonian}

We assume that all atoms are initially in the state $|a\rangle$ with 
total angular momentum zero, and that this state is Raman-coupled 
to the state $|b\rangle$ with a winding number 
$l$ by using a pair of LG laser fields, one with a winding number 
zero and the other with winding number $l$. We further assume that 
two-body interactions are weak enough, and the external harmonic 
trapping frequencies $\omega_{r,j}=\omega_r$ and 
$\omega_{z,j}=\omega_z$ are large enough, that the mode profiles 
of the two atomic states are not significantly modified from their 
single particle forms $\psi_{0,l}(\bm{r})$~\cite{97MCWW,98SC}.

Under these assumptions the matter-wave field operator may be expressed as 
\bq\label{field_operator}
\hat{\Psi}(\bm{r},t)=\hat{a}(t)\psi_0(\bm{r})+\hat{b}(t)\psi_l(\bm{r}),
\eq 
where $\hat{a}$ and $\hat{b}$ are the bosonic annihilation 
operators for the states $|a\rangle$ and $|b\rangle$, and the 
positive frequency components of the Rabi frequencies are
\bq
    \Omega_1^{(+)}(\bm{r})&=&\Omega_1\phi_l(\bm{r})\ ,\quad
    \Omega_2^{(+)}(\bm{r})=\Omega_2\phi_0(\bm{r}).
    \label{two-mode2}
    \eq
When the LG beam remains well collimated over the longitudinal 
extent of the condensate, $d_z \ll k_0w^2/2$, the laser mode 
functions $\phi_{0,l}$ at the beam waist and the condensate mode functions $\psi_{0,l}$ may be expressed as
    \bq\label{2mode}
    \phi_l(\bm{r})&=&\frac{1}{\sqrt{l!\pi}}\left(\frac{r}{w_0}\right)
    ^le^{-r^2/2w_0^2}e^{il\varphi}e^{ikz},\\
    \psi_l(\bm{r})&=&\frac{1}{\sqrt{l!\pi d_r^2}}\left(\frac{r}{d_r}\right)
    ^le^{-r^2/2d_r^2}e^{il\varphi}
    \psi_z(z).
    \eq
We assume that $l>0$, and that the radial and longitudinal atomic 
oscillator ground state widths are the same for both atomic states 
$d_r = \sqrt{\hbar/(M\omega_r)}$, and $d_z =\sqrt{\hbar/(M\omega_z)}$~\cite{oscillator_length}.  
We further assume that the longitudinal trapping is much stronger than the 
radial trapping, $d_r>d_z$, and that the longitudinal wave 
function remains fixed as the ground state 
$\psi_z(z)=e^{-z^2/2d_z^2}/\sqrt{\pi^{1/2} d_z}$. The two-mode 
approximation is then valid under the conditions $Na_{ij} \lesssim 
d_{k}$~\cite{97MCWW}, where $i,j=a,b$ and $k=r,z$.

Substituting Eqs.~(\ref{field_operator})-(\ref{2mode}) into the 
system Hamiltonian~(\ref{HHH}) with $E_j=\int d^3 r \psi^*_j\hat{H}_0\psi_j$, 
and transforming to a rotating frame using the 
unitary transformation 
$U_0=\exp\{-i [E_a(\hat{a}^{\dagger}\hat{a}+\hat{b}^{\dagger}\hat{b})/{\hbar}
-(\omega_2-\omega_1)\hat{b}^{\dagger}\hat{b} ]t\}$, finally yields the reduced two-mode Hamiltonian
    \bq\label{2modeHamiltonian}
    \hat{H}&=&\hbar\omega_r\left[(l+\delta/\omega_r)\hat{b}^{\dagger}\hat{b} +G_{aa}
    \hat{a}^{\dagger}\hat{a}^{\dagger}\hat{a}\hat{a}
    +G_{bb} \hat{b}^{\dagger}\hat{b}^{\dagger}\hat{b}\hat{b}\right.\nonumber\\
    &\ &\qquad\quad  + \left.2 G_{ab} \hat{a}^{\dagger}\hat{b}^{\dagger}\hat{b}\hat{a} -
    g(\hat{a}^{\dagger}\hat{b} + \hat{a}\hat{b}^{\dagger})\right],
    \eq
where
we have defined $\delta=\omega_2-\omega_1+(E_b-E_a)/\hbar$ as the 
two-photon detuning, which may be controlled experimentally. 
The dimensionless coupling constants for the atom-atom and 
atom-light interactions are given by
    \bq
    G_{jj'}&=&\frac{\chi_{jj'}V_{jj'j'j}}{2 \hbar\omega_r}\quad (j,j'=a,b)\label{defG},\\
    g&=&\frac{2\Omega_1\Omega_2 U_{0ll0}}{\omega_r\Delta}, \label{coeffs}
    \eq
with
    \bq
    V_{klmn} &=& \int d^3r \psi_k^* \psi_l^* \psi_m \psi_n\nonumber\\
    &=&\frac{1}{2\pi d_r^2\sqrt{2\pi d_z^2}}\frac{(k+l)!
    \delta_{k+l=m+n}}{2^{k+l}\sqrt{k!l!m!(k+l-m)!}}\label{defV},\\
    U_{klmn} &=& \int d^3r \psi_k^* \phi_l^* \psi_m \phi_n\nonumber\\
    &=&\frac{\zeta^{k+2l-m}}{\pi (\zeta^2+1)^{k+l+1}}\frac{(k+l)!
    \delta_{k+l=m+n}}{\sqrt{k!l!m!(k+l-m)!}}.
    \eq
The parameter $\zeta\equiv d_r/w_0$ characterizes the ratio of the 
radial oscillator ground state width to the focused light spot 
size, and the Kronecker delta enforces angular-momentum 
conservation for the atom-field interactions. Henceforth 
the energies, lengths, and angular momenta are measured in 
units of $\hbar\omega_r$, $d_r$, and $\hbar$, respectively.

%
\subsection{Angular-momentum representation}

In order to study the quantum dynamics described by the 
Hamiltonian~(\ref{2modeHamiltonian}), we introduce the following 
operator combinations that map our problem to Schwinger's 
angular-momentum representation~\cite{97MCWW},
    \bq
    \hat{J}_+ &=&
    \hat{a}\hat{b}^{\dagger},\qquad
    \hat{J}_-=\hat{a}^{\dagger}\hat{b},\\
    \hat{J}_x &=& \frac{1}{2}(\hat{J}_++\hat{J}_-),\\
    \hat{J}_y &=& \frac{1}{2i}(\hat{J}_+-\hat{J}_-),\\
    \hat{J}_z &=&
    \frac{1}{2}(\hat{b}^{\dagger}\hat{b}-\hat{a}^{\dagger}\hat{a}).
    \eq
These operators obey the SU(2) commutation relations 
$[\hat{J}_i,\hat{J}_j]=i \epsilon_{ijk}\hat{J}_{k}$ where 
$\epsilon_{ijk}$ denotes the Levi-Civita antisymmetric symbol, and 
the Casimir invariant is found to be $\hat{J}^2=\hat{J}_x^2 + 
\hat{J}_y^2 + \hat{J}_z^2=j\left(j+1\right)$ with $j=N/2$. For the 
basis $|j, m\rangle$ with $m=-j,-j+1,\dots, j$, which are 
eigenstates of $\hat{J}_z$, the operators satisfy
    \bq
    \hat{J}_z|j,m\rangle&=&m|j,m\rangle,\\
    \hat{J}^2|j,m\rangle&=&j(j+1)|j,m\rangle,\\
    \hat{J}_{\pm}|j,m\rangle&=&\sqrt{(j\pm m+1)(j\mp m)}|j,m\rangle.
    \eq
In the angular-momentum representation the Hamiltonian 
(\ref{2modeHamiltonian}) then becomes
\bq \hat{H}=h_z\hat{J}_z+
\chi\hat{J}_z^2 - 2 g \hat{J}_x,
\eq
where
\bq
    h_z &=& l+\delta - (N-1)\chi_-,\\
    \chi_- &=& G_{aa} - G_{bb}, \quad
    \chi = G_{aa} + G_{bb} - 2 G_{ab}.
\eq
The angular basis states $|j,m\rangle$ and the Fock basis states 
$|n_a, n_b\rangle_{\rm F}$ are related by $|j,m\rangle =|N/2-m,N/2+m\rangle_{\rm F}$.


\section{Quantum dynamics}\label{results}

In this section we explore the quantum dynamics of the two-mode 
model, first with no atomic collisions, and then we study the 
collapse and revivals that occur with atomic collisions.  In 
particular, we highlight that the collapse and revivals may be 
monitored by following the motion of quantized vortices that 
appear in the spatial density profile, and we shall also study 
the resulting atom statistics.


\subsection{No atomic collisions}

In the absence of atomic collisions, $G_{ij}=0$, giving 
$h_z=l+\delta$ and $\chi=0$, the two-mode Hamiltonian reduces to 
the form of two simple coupled harmonic oscillators. Then the 
solution of the Heisenberg equations of motions for $\hat{a}(t)$ 
and $\hat{b}(t)$ is found to be 
\bq \left[
\begin{array}{c}
\hat{a}(t)\\
\hat{b}(t)
\end{array}
\right]
=e^{-i(l+\delta)/2}
\left[
\begin{array}{cc}
{\cal A}(t) & {\cal B}(t)\\
{\cal B}(t) & {\cal A}^*(t)
\end{array}
\right]
\left[
\begin{array}{c}
\hat{a}(0)\\
\hat{b}(0)
\end{array}
\right],
\eq
where
    \bq
    {\cal A}(t) &=& \cos \nu t + \frac{i(l+\delta)}{2\nu}\sin \nu t,\\
    {\cal B}(t) &=& \frac{ig}{\nu}\sin \nu t,
    \eq
and the effective Rabi frequency $\nu$ is defined as 
\bq 
\nu=\frac{\sqrt{(2g)^2+(l+\delta)^2}}{2}. 
\eq

We consider an initial state where all atoms are condensed in the 
hyperfine state $|a\rangle$ with zero OAM. 
If we take a number conserving initial state with definite atom number, 
$|\Psi(0)\rangle_N=|N,0\rangle_{\rm F}=|N/2,-N/2\rangle$, then the half 
population difference $\hat{J}_z=(\hat{b}^{\dagger}\hat{b}-\hat{a}^{\dagger}\hat{a})/2$ 
evolves in time as
    \bq\label{WF}
    _N\langle \Psi
    (0)|\hat{J}_z|\Psi (0)\rangle_N = -N\frac{(l+\delta)^2 + (2g)^2
    \cos (2\nu t)}{2(2\nu)^2}.
    \eq
The total-angular-momentum operator is just proportional to the 
number operator of the state $|b\rangle$, 
\bq
\hat{L}=l\hat{b}^{\dagger}\hat{b} = l (\hat{J}+\hat{J}_z), 
\eq 
and the fluctuations in angular momentum coincide with the number 
fluctuations of the state $|b \rangle$. The time evolution of the 
angular momentum for an initial Fock state is given by
\bq\label{LF}
_N\langle \Psi(0) | \hat{L} | \Psi(0) \rangle_N  =
\frac{Nl\sin^2(\nu t) }{1+(l+\delta)^2/(2g)^2}.
\eq
We note that the maximum population transfer occurs at the resonant 
two-photon detuning $\delta=-l$, and the parameter $g$ in 
Eq.~(\ref{coeffs}) describing the single-particle Raman coupling 
between the LG field and the BECs decreases with increasing winding number $l$.

We now repeat the same calculation, but using a Glauber coherent 
state as an example of an initial BEC displaying BBS
\bq
|\Psi(0)\rangle_{\alpha}
&=&|\alpha,0\rangle=\sum_{N=0}^{\infty}C_{\alpha,N}|N,0\rangle_{\rm
F}, \eq where \bq C_{\alpha,N}=
e^{-|\alpha|^2/2}\frac{\alpha^N}{\sqrt{N!}},
\eq 
with $|\alpha|^2\equiv \bar{N}$ being the average number of atoms. 
The number and phase fluctuations of the coherent state are given by 
$\Delta n = |\alpha|$, and $\Delta \phi \simeq 1/(2|\alpha|)$, 
respectively. The time evolutions of the half population 
difference $_{\alpha}\!\langle \Psi (0)|\hat{J}_z|\Psi (0)\rangle_{\alpha}$ 
and the total angular momentum 
$_{\alpha}\!\langle \Psi (0)|\hat{L}|\Psi (0)\rangle_{\alpha}$ are 
easily shown to coincide with Eqs. (\ref{WF}) and (\ref{LF}) with 
the replacement $N \to \bar{N}$. In other words, in the absence of 
the atomic collisions, the physical quantities are therefore 
independent of whether the state of the initial BEC exhibits BBS or not.


\subsection{Collapse and revivals}

The two-mode Hamiltonian exhibits a rich variety of quantum 
dynamics depending on the relative magnitude of the atomic 
collisions and atom-field coupling, and the OAM of the LG beam. 
The system dynamics can be categorized into three typical 
parameter regimes~\cite{L}, namely 
(i)~the Rabi regime $g \gg \chi N$, 
(ii)~the Josephson regime $g \ll \chi N \ll g N^2$, and 
(iii)~the Fock regime $gN \ll \chi $. 
Oscillations between the two BEC states 
are almost perfect in the Rabi regime at the resonance condition 
$h_z=0$, but the transfer between the coupled BECs reduces as the 
Josephson regime is approached.  In the Fock regime, the Rabi 
oscillations are suppressed and the number in each internal state 
is almost a constant, like the case of self-trapping in a 
double-well potential or the Mott state in optical lattices. The 
critical point of the self-trapping transition is $\chi N \simeq 
4g$, which lies in the Josephson regime~(ii). Here we concentrate on 
the situation where the system is in the Rabi regime, considering 
for concreteness the parameter values $N=\bar{N}=100$, $\chi=5\times 
10^{-3}$, and $g=10$, corresponding to a condensate with a 
mesoscopic total number of atoms. In the following numerical study we choose 
$l=2$ as a representative example, and take the two-photon detuning 
as $\delta=-l+(N-1)\chi_-$, so that $h_z=0$. With these 
parameters we anticipate almost complete Rabi oscillations of the 
population difference between the two condensate components.

To proceed we introduce two initial states that represent NC and 
BBS states for the initial BEC.  First, for the NC state we choose 
a state with a well-defined atom number $|N,0\rangle_{\rm F}$ in 
Fock-space representation. The subsequent quantum dynamics of the 
coupled BECs may then be represented in the angular-momentum basis as
    \bq\label{Fock}
    |\Psi(t)\rangle_N=\sum_{m=-j}^j
    A_m(t)|j,m\rangle,
    \eq
where $j=N/2$, and the normalization is given by $\sum_m |A_m(t)|^2=1$.  
On the other hand, for an initial condition with 
BBS where the initial state is a coherent state 
$|\alpha,0\rangle$, the time evolution of the coupled BECs may be 
represented in the angular-momentum basis as
    \bq\label{coherent}
    |\Psi(t)\rangle_{\alpha}&=&\sum_{N=0}^{\infty}
    C_{\alpha,N}\sum_{m=-j'}^{j'}
    A_{m}^{(j')}(t)\delta_{N,2j'}|j',m\rangle,
    \eq
where there is no dynamical coupling between subspaces with 
different total number of atoms since the Hamiltonian conserves 
the atom number~\cite{07CMM}.

\begin{figure}[t]
\begin{center}
\epsfig{file=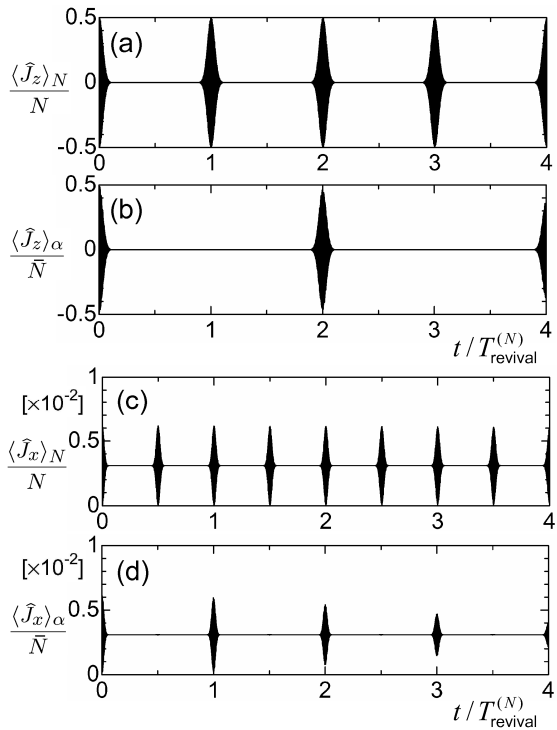,width=7.5cm}
\caption{
Time evolutions of 
(a) $\langle \hat{J}_z \rangle_N$, 
(b) $\langle \hat{J}_z \rangle_{\alpha}$, 
(c) $\langle \hat{J}_x \rangle_N$, 
(d) $\langle \hat{J}_x \rangle_{\alpha}$ 
for $N=\bar{N}=100$, $\chi = 5\times 10^{-3}$, and $g=10$.
}\label{CandR_J}
\end{center}
\end{figure}

Figure~\ref{CandR_J} shows the numerically calculated time 
evolution of $\langle \hat{J}_{x}\rangle_N$ and 
$\langle \hat{J}_{z} \rangle_N$ for an initial Fock state, 
and of  $\langle\hat{J}_{x}\rangle_{\alpha}$ and 
$\langle\hat{J}_{z}\rangle_{\alpha}$ for an initial coherent state. 
In both cases atomic collisions cause the anticipated complete Rabi 
oscillations between the two BECs to be modulated by a sequence of 
collapses and revivals, the revivals being a characteristic 
consequence of the granular nature of the matter wave fields. The 
components $\hat{J}_{x, y}$ characterize the coherence between the 
states $|a\rangle$ and $|b\rangle$, while $\hat{J}_z$ corresponds 
to half their population difference.  We note that $\langle 
\hat{J}_x \rangle$ exhibits a revival in the middle of 
the collapse of $\langle \hat{J}_z \rangle$ for both coherent and 
Fock initial states.

There is a significant difference between the period of revivals 
for the NC and BBS initial states. In the Rabi regime, and for the 
resonant case $h_z=0$, the collapse time and revival time in the 
population difference $\langle \hat{J}_z \rangle$ are numerically 
found to be
\bq
T_{\rm collapse}^{(N)} = \frac{\cal C}{\chi\sqrt{N}},
\qquad T_{\rm revival}^{(N)} = \frac{2\pi}{\chi}
\eq
for the initial Fock state, and
\bq T_{\rm collapse}^{(\alpha)}\simeq
T_{\rm collapse}^{(N)},\qquad T_{\rm revival}^{(\alpha)} = 2T_{\rm
revival}^{(N)}
\eq
for the initial coherent state, respectively, 
where the constant is ${\cal C}\simeq 10$. This factor of 2 difference 
in the revival period is also found in the double-well system and 
for the Ramsey fringe experiment proposed by Search, and the 
revival period agrees with that he found for the resonant case~\cite{01S}.

For the sake of completeness we comment briefly on the effect on 
our findings of selecting a nonzero value of the parameter $h_z$. 
For the case $h_z \ne 0$, which would be possible using a 
Feshbach resonance, the system enters the Josephson regime as 
$\chi_-$ increases, and this reduces the collapse time for both 
the Fock and coherent states. The Rabi oscillations are then no 
longer perfect, and the coherent state collapses much faster than 
the Fock state. Although the revival time becomes longer, the 
factor of two difference between Fock and coherent initial states 
remains. This is different from the situation in the Ramsey 
technique, where the system evolves freely in time (without 
coupling between the BECs) between the Ramsey pulses and the 
coupling term proportional to $\hat{J}_x$ in the Hamiltonian is 
zero. In the present case, by contrast, the Raman coupling is on at 
all times.

Finally we emphasize that the factor of two difference in revival 
period depends only on whether the initial state is NC or not.  For 
instance, if the initial state is chosen as number-conserving 
coherent state with binomial statistics~\cite{72ACGT}, then the 
period of the revival will still be half that of the Glauber coherent state.


\subsection{Quantized vortex motion}

\begin{figure*}[t]
\includegraphics[scale=0.9]{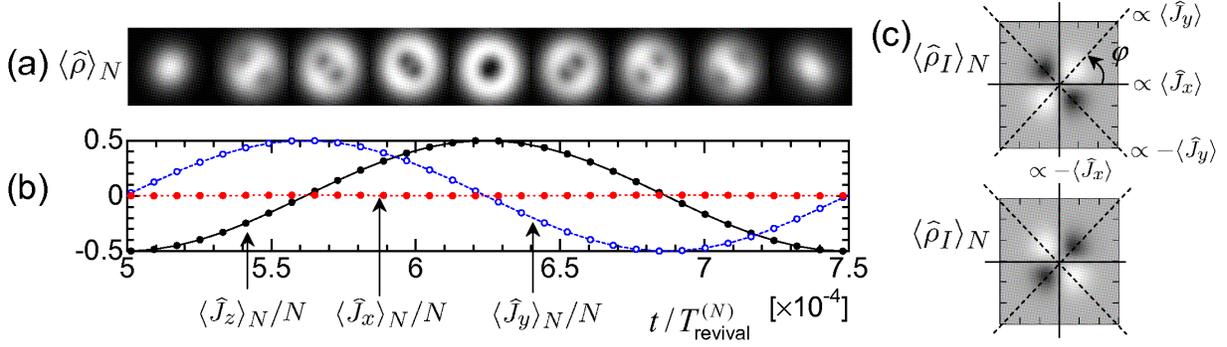}
\caption{ 
(a) Time sequence of density profiles $\langle 
\rho(\bm{r},t)\rangle_N$ given by Eq.~(\ref{density}), and (b) 
expectation values of the pseudospin components during one Rabi 
cycle. Solid curve with the filled dots shows $\langle 
\hat{J}_z\rangle_N$, dotted curve with the open dots $\langle 
\hat{J}_y\rangle_N$, and dotted curve with the filled dots 
$\langle \hat{J}_x\rangle_N$, respectively. (c) Off-diagonal part 
of the density $\langle \rho_I(\bm{r},t)\rangle_N$ given by 
Eq.~(\ref{OD}). Upper and lower panels show the typical patterns 
for the first half and latter half of a Rabi cycle, respectively. 
The alignment of the two density maxima (mimina) rotates by 
$\pi/2$ anytime when the $\langle \hat{J}_y \rangle_N$ crosses the 
zero. 
}\label{fig_one_cycle}
\end{figure*}

The results of the previous section show that the collapse and 
revivals in the dynamics of the quantum vortex coupler can be used 
to test whether the initial state of the BEC is described as a NC 
state or a BBS state, and these results are in perfect accord with 
previous works on this topic.  However, measuring the components 
of the angular momentum in Schwinger's representation raises issues 
of how to detect these quantities.  Here we propose to detect the 
collapses and revivals by monitoring the spatial density profile 
of the Raman-coupled BECs, and in particular, the motion of the 
quantized vortices that appear in the profile.  Collapse and 
revival dynamics have previously been observed by monitoring the 
spatial density profile in a two-component $^{87}$Rb 
BEC~\cite{99JILA}, and a Raman vortex coupler using LG beams has 
now been realized experimentally~\cite{06NIST}. These experiments 
were performed for large atom numbers where the granular nature of 
the matter wave fields is not relevant, but they show that the proposed approach has some validity.

To proceed we investigate how the quantum dynamics can manifest 
itself in the spatial density profile of the coupled BECs. From 
Eq.~(\ref{field_operator}), the density operator 
$\hat{\rho}(\bm{r},t)$ may be written as
\bq\label{density}
\hat{\rho}(\bm{r},t)=\hat{\Psi}^{\dagger}(\bm{r},t)\hat{\Psi}(\bm{r},t)
=\hat{\rho}_0(\bm{r},t)+ \hat{\rho}_{\rm I}(\bm{r},t),
\eq
where
\bq
\hat{\rho}_0(\bm{r},t)\!&=&\!|\psi_0(\bm{r})|^2
\left[\left(1+\frac{r^{2l}}{l!}\right)\hat{J}(t)-
\left(1-\frac{r^{2l}}{l!}\right)\hat{J}_z(t)\right],\nonumber\\
&\ &\\
\hat{\rho}_{\rm I}(\bm{r},t)\!&=&\!\frac{r^l|\psi_0(\bm{r})|^2}{\sqrt{l!}}
\left[\hat{J}_-(t) e^{il\varphi}+\hat{J}_+(t) e^{-il\varphi}\right]\nonumber\\
&=&\!\frac{r^l|\psi_0(\bm{r})|^2}{\sqrt{l!}}\left[\hat{J}_x(t)
\cos(l\varphi)+\hat{J}_y(t) \sin(l\varphi)\right], \label{OD}
\eq
and the total density profile for both atomic states is given by 
$\langle \hat{\rho}(\bm{r},t) \rangle=\langle 
\hat{\rho}_0(\bm{r},t) \rangle+\langle \hat{\rho}_{\rm I}(\bm{r},t) \rangle$.

From this equation we see that the spatial density profile is 
directly related to the pseudospin components, hence it will 
also display the collapse and revival phenomenon. 
Since the qualitative properties of the density profile are the 
same for both the Fock and coherent initial states, with the 
caveat that the revival period is twice as long for the coherent 
state, here we show results for the case of an initial number state.

Figure~\ref{fig_one_cycle}(a) shows the evolution of the density 
profile $\langle \hat{\rho}(\bm{r},t)\rangle_N$ over one Rabi 
oscillation for times sufficiently short that the first collapse 
is barely noticeable, and Fig.~\ref{fig_one_cycle}(b) shows the 
corresponding evolution of $\langle \hat{J}_i 
\rangle_N/N,i=x,y,z$ for the number state 
here times are scaled to the revival time $T_{\rm revival}^{(N)}$. 
For the earliest time $t/T_{\rm revival}^{(N)}=5\times 10^{-4}$ in these plots, 
$\langle \hat{J}_z \rangle_N/N \simeq -1/2$, meaning that the 
condensate is almost entirely in the nonrotating Gaussian mode, 
and the left-most gray-scale density profile in 
Fig.~\ref{fig_one_cycle}(a) shows this Gaussian density profile. 
As time progresses, however, the Raman coupling transfers more 
population to the rotating BEC mode with $l=2$, and the second 
through fourth density profiles in Fig.~\ref{fig_one_cycle}(a) 
show this process in which two vortex cores with winding number 
$l=1$ progressively make their way towards the center of the 
condensate. Halfway through the Rabi oscillation the BEC is 
composed purely of the rotating state, corresponding to the time 
$t/T_{\rm revival}^{(N)}=6.25\times 10^{-4}$ in 
Fig.~\ref{fig_one_cycle}(b) where $\langle \hat{J}_z \rangle_N/N 
\simeq 1/2$, and the density profile is now that for a pure 
matter-wave vortex with $l=2$ centered at the origin.  For longer 
times the process reverses up until $t/T_{\rm 
revival}^{(N)}=7.5\times 10^{-4}$ when the next Rabi oscillation starts.

During a single Rabi oscillation the distance between the two 
$l=1$ matter-wave vortices is related to the population difference 
$\langle \hat{J}_z\rangle_N$.  The angle of the line joining the 
two $l=1$ vortex cores (where the density goes to zero) may be 
understood by looking at the off-diagonal part of the density 
profile $\langle \rho_I(\bm{r},t)\rangle_N$. As seen from 
Eq.~(\ref{OD}), the angle of the line joining the vortex cores is 
determined by the pseudospin components $\langle \hat{J}_x 
\rangle$ and $\langle\hat{J}_y \rangle$. For example, the spatial 
distribution of the off-diagonal part of the density profile at 
the angle $\varphi=0$ is always proportional to $\langle \hat{J}_x 
\rangle$, whereas at the angle $\varphi=\pi/4$ it is proportional 
to $\langle \hat{J}_y \rangle$. For the parameters of this example 
we have that $|\langle \hat{J}_x \rangle_N| \ll |\langle 
\hat{J}_y\rangle_N|$, see Fig.~\ref{fig_one_cycle}(b). Then, 
viewing the expectation values of the pseudospin components as 
parameters on a Bloch sphere, one Rabi oscillation corresponds to 
a round trip between the south pole ($\langle \hat{J}_z \rangle_N/N=-1/2$) 
and the north pole ($\langle \hat{J}_z \rangle _N/N=1/2$), 
with the shortest path being taken on the Bloch sphere. Since the 
component $\langle \hat{J}_x\rangle$ is small compared to the 
other two components, the off-diagonal term is approximately given 
by $\langle \rho_{I} \rangle \simeq 
r^l|\psi_0(\bm{r})|^2\langle \hat{J}_y(t) \rangle \sin(l\varphi)/\sqrt{l!}$, 
which illustrates the change in profile in Fig.~\ref{fig_one_cycle}(c) 
for times before and after the system evolves halfway through 
the Rabi oscillation in Fig.~\ref{fig_one_cycle}(b). 
When the value $\langle \hat{J}_y \rangle_N$ crosses zero halfway 
through the Rabi oscillation at 
$t/T_{\rm revival}^{(N)}=6.25\times 10^{-4}$ in 
Fig.~\ref{fig_one_cycle}(b), the positions of the off-diagonal 
density maxima and minima are interchanged. This gives rise to the 
change in the angle of the line joining the two $l=1$ vortex cores 
in Fig.~\ref{fig_one_cycle}(a).  We remark that if the dynamics 
takes the longer path on the Bloch sphere, i.e., as happens when 
$\hat{J}_x$ and $\hat{J}_y$ become comparable and exchange their 
values, the position of the vortices can rotate around the 
condensate. Hence the radius and angle of the vortex position 
generally follow a trajectory on the Bloch sphere.

\begin{figure}
\begin{center}
\epsfig{file=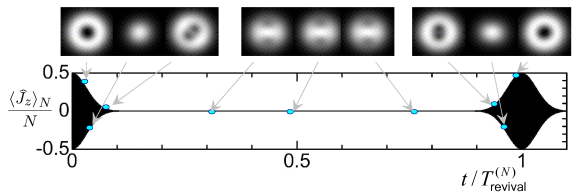,width=7.5cm}
\caption{ 
Density profiles $\langle \rho(\bm{r},t)\rangle_N$ 
(upper panels) and half population difference $\langle \hat{J}_z\rangle_N/N$ 
(lower panel) for 
times allowing for collapse and revivals. The density plots are 
shown at the times indicated by small circles in the lower panel. 
}\label{fig_rev}
\end{center}
\end{figure}

Figure~\ref{fig_rev} shows a time sequence of density plots over 
time scales long enough to allow for a collapse and revival, the 
lower plot showing the evolution of $\langle \hat{J}_z\rangle_N/N$ 
versus time.  For short times the system undergoes Rabi 
oscillations as in Fig.~\ref{fig_one_cycle}, but modulated by a 
collapse envelope are shown for times $t/T_{\rm revival}^{(N)}<0.1$, 
in the upper left panel. In the collapse region 
$0.1<t/T_{\rm revival}^{(N)}<0.9$ when $\langle 
\hat{J}_z\rangle_N/N\rightarrow 0$, the density profile is to all 
intents and purposes stationary and the vortex cores assume fixed 
positions, see the central upper panel. Finally, for 
$t/T_{\rm revival}^{(N)}>0.9$ the first revival starts, the vortex 
cores get back into motion, and the Rabi oscillations start 
again, see the right most upper panel.  Thus the motion of the 
vortex cores in the density profile provides a convenient means to 
monitor collapse and revivals in the quantum dynamics of the 
matter wave vortex coupler. Density measurements thus represent a 
powerful tool to measure the dynamics of the collapse and revivals 
as well as the motion of the condensate on the Bloch sphere, and 
as such allow one to measure the factor of two difference in the 
revival times for initial state displaying Bose broken symmetry or 
being number conserving.


\subsection{Atom statistics}

\begin{figure}[t]
\begin{center}
\epsfig{file=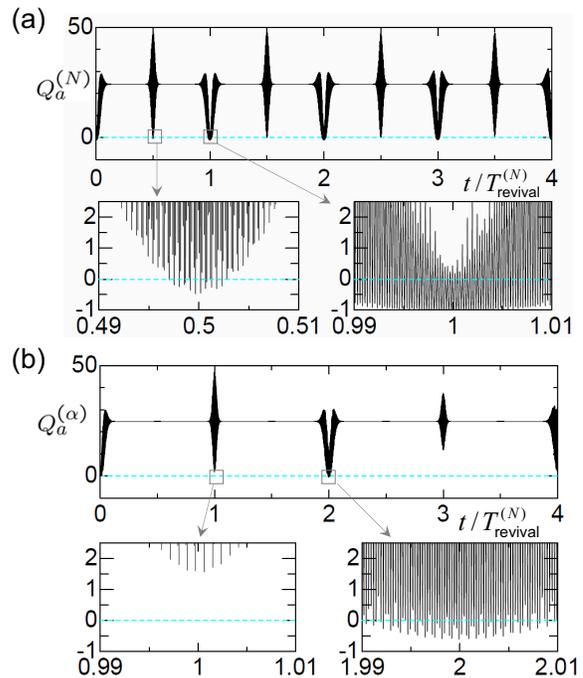,width=7.5cm}
\caption{
Mandel's $Q$ parameter defined by Eq.~(\ref{MandelQ}) 
for initial (a) Fock and (b) coherent state. 
The value becomes $-1$ near the revival for the initial 
Fock state, while it is always larger than $-1$. 
}\label{CandR_Q}
\end{center}
\end{figure}

For completeness we now briefly discuss the atom statistics of the 
system as it undergoes a series of collapses and revivals. A 
familiar measure of the atom-number fluctuations is Mandel's 
$Q$ parameter~\cite{Qparameter} 
    \bq\label{MandelQ} Q_j=\frac{\langle(\Delta
    \hat{n}_j)^2\rangle-\langle\hat{n}_j\rangle}{\langle \hat{n}_j
    \rangle}, \qquad j=a,b,
    \eq
where $Q=0$ corresponds to a Poissonian,  $-1 \le Q < 0$ to a 
sub-Poissonian, and $Q>0$ to a super-Poissonian  distribution. 
In particular, $Q=-1$ corresponds to a Fock state with $\langle 
\Delta n_j\rangle=0$. As expected and illustrated in 
Fig.~\ref{CandR_Q}, the Mandel parameter also exhibits the factor of 
two difference in the revival times depending on the U(1) symmetry 
of the condensate. For the initial Fock state [Fig.~\ref{CandR_Q} 
(a)], $Q_a=-1$ initially, and $Q_a(t)$ returns to that value 
periodically near the revivals. For an initial coherent state, on 
the other hand, the Mandel parameter is initially $Q_a=0$. It then 
becomes negative near the revivals, indicative of a 
sub-Poissonian distribution, but never reaches the Fock state 
value $Q_a=-1$. Note also that the average values of $Q_a$ for the 
coherent and Fock state approach each other as the number of atoms increases.


\subsection{Effects of thermal dissipations}\label{exp}

To assess the feasibility of an experimental demonstration of 
these predictions, we consider the specific example of a  $^{87}$Rb 
condensate with Raman coupling between the two hyperfine states 
$|a\rangle=|F=1,m_F=-1\rangle$ and $|b\rangle=|F=1,m_F=1\rangle$. 
The trapping potentials of these two states are almost identical, 
$V_1 (\bm{r})\simeq V_2(\bm{r})$. It is experimentally clearly 
desirable to have a relatively short revival time, so that a tight 
condensate confinement is preferable. Taking $\omega_r = 2\pi 
\times 100$ Hz and $\omega_z = 2\pi \times 300$ Hz, the 
corresponding oscillator lengths are then $d_r \simeq  1$ $\mu$m, 
$d_z\simeq 0.6$ $\mu$m. All $s$-wave scattering lengths are 
approximately the same, $a_{aa}\simeq a_{bb}\simeq a_{ab} \simeq 
5.5$ nm. For this set of parameters, the validity of the two-mode 
model requires that the total number of atoms has to be $N 
\lesssim 200$. For the case of an OAM $l=1$, these values result 
in the dimensionless parameters 
$G_{aa}=2G_{bb}=2G_{ab}\simeq 3.7\times 10^{-3}$ and thus 
$\chi=G_{aa}/2=1.8 \times 10^{-3}$ 
[see Eqs.~(\ref{defG}) and (\ref{defV})]. Typical time scales are 
then estimated as $T_{\rm Rabi} \simeq 1.4 \times 10^{-3}\ {\rm 
s}$, $T_{\rm collapse}^{(N)} \simeq 0.5\ {\rm s}$, 
$T_{\rm revival}^{(N)} \simeq  5.5\ {\rm s}$, and 
$T_{\rm revival}^{(\alpha)}\simeq  11.0\ {\rm s}$, respectively. 
Hence the difference in the revival periods of the Fock and coherent states 
should be observable in a condensate with a relatively long lifetime on 
the order of a few tens of second.

We note that there is a caveat with respect to our proposal in that 
it involves multiply-charged matter-wave vortices with winding numbers 
whose magnitude is greater than one.  In particular, it is well known 
that if a multiply charged vortex is initiated and allowed to evolve 
freely it will quickly decay into multiple singly charged vortices in the presence 
of even a small background component of thermal atoms~\cite{04MIT}. 
The decay time $\tau_d$ for a doubly-quantized vortex ranges from one to a few tens 
of milliseconds, depending on the total number of atoms.  

However, this does not pose a problem for the Raman coupler analyzed here, as the 
matter-wave vortices are not freely moving, but rather are externally driven by the LG beams.  
More specifically, the applied LG laser fields continually drive the matter-wave field 
between the zero-charged ($l=0$) and doubly charged ($l=2$) vortex states with 
a time period corresponding to the inverse Rabi frequency 
$T_{\rm Rabi} \simeq 1.4 \times 10^{-3}\ {\rm s}$.  (This is true even in the collapse region 
where the time-independent density profile arises from an average over many oscillating 
components with slightly different periods.)  Thus, as long as $T_{\rm Rabi} \ll \tau_d$ 
there will be negligible decay of the doubly charged vortex created in any Rabi cycle 
before it is returned to the zero-charged state and the next Rabi cycle starts. 


\section{Summary}\label{summary}

\begin{figure}[t]
\begin{center}
\epsfig{file=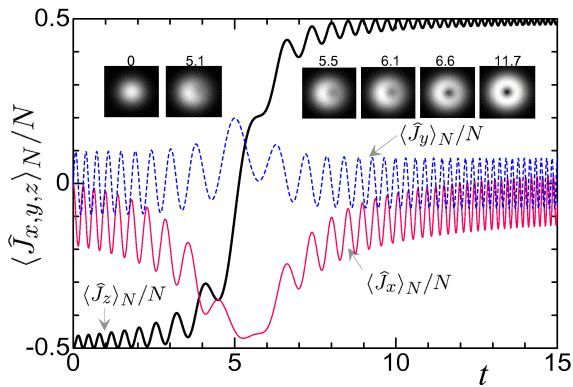,width=7.5cm}
\caption{
Population transfer from the state $|a\rangle$ to 
$|b\rangle$ with the linear sweep of the two-photon detuning. 
The insets and the numbers express the intensity of 
the interference pattern and the corresponding (dimensionless) time. 
}\label{fig_transfer}
\end{center}
\end{figure}

In summary, we have studied the quantum dynamics of the vortex 
coupler using the Laguerre-Gaussian beams. As a consequence of the 
quantization of the matter-wave field, the atoms are found to 
undergo a series of collapses and revivals whose period is directly 
observable as an off-axis motion of the quantized vortex cores that appear in the condensate 
density. The angle of the vortex also provides direct information 
on its path on the Bloch sphere that describes its dynamics in the 
Schwinger representation. An important feature is that the 
characteristic time scale of the collapse and revivals differs by 
the factor 2 depending on whether the description of the 
condensate is in terms of a number-conserving state or a state 
with broken U(1) symmetry~\cite{97WCWTW,Walls2,01S}.

We finally note that instead of generating the collapse and revivals 
used to test the appropriate theoretical description of the
condensate, our scheme can also be modified slightly to 
permanently transfer the condensate population to the state 
$|b\rangle$. All that is required is to sweep the two-photon 
detuning linearly in time at an appropriate rate. 
We illustrate this technique with the parameters $N=100$, 
$\chi = 5\times 10^{-3}$, $g=2$, and $\delta = 20 - 4 t$ in Fig.~\ref{fig_transfer}. 
Since both the initial and final states are in the Josephson regime, no 
detectable collapse and revival occurs in any observables. The 
$\hat{J}_x$ and $\hat{J}_y$ components have maximum values when 
the detuning $\delta$ crosses the resonance, and a significant 
change in $\hat{J}_z$ occurs at this point. Note that a 
population transfer is achievable independently of the U(1) 
symmetry, although the variances and atom statistics differ significantly in both cases.


\begin{acknowledgments}
We are thankful to Dr.~D.~Meiser and Dr.~B.~P.~Anderson for fruitful 
discussions and comments. This work is supported in part by the 
US Office of Naval Research, by the National Science Foundation, 
by the US Army Research Office, and by the National Aeronautics and Space Administration.
\end{acknowledgments}



\begin{thebibliography}{50}
\bibitem{oscillator_length}
Strictly speaking, the frequencies of the harmonic potentials and thus the oscillator lengths
for the atomic states $|a\rangle$ and $|b\rangle$ are different.
However, this difference can be erased by adjusting
(i) the laser frequency so that the $\omega_r^{(a)}=\omega_r^{(b)}$, or
(ii) the coupling constant of the interatomic interaction
so that $d_r^{(a)}=d_r^{(b)}$.
\end{thebibliography}
\end{document}